# Mixed Quality of Service in Cell-Free Massive MIMO

Manijeh Bashar, *Student Member, IEEE*, Kanapathippillai Cumanan, *Member, IEEE*, Alister G. Burr, *Member, IEEE*, Hien Quoc Ngo, *Member, IEEE*, and H. Vincent Poor, *Fellow, IEEE*

*Abstract*—Cell-free massive multiple-input multiple-output (MIMO) is a potential key technology for fifth generation wireless communication networks. A mixed quality-of-service (QoS) problem is investigated in the uplink of a cell-free massive MIMO system where the minimum rate of non-real time users is maximized with per user power constraints whilst the rate of the real-time users (RTUs) meet their target rates. First an approximated uplink user rate is derived based on available channel statistics. Next, the original mixed QoS problem is formulated in terms of receiver filter coefficients and user power allocations which can iteratively be solved through two sub-problems, namely, receiver filter coefficient design and power allocation, which are dealt with using a generalized eigenvalue problem and geometric programming, respectively. Numerical results show that with the proposed scheme, while the rates of RTUs meet the QoS constraints, the 90%-likely throughput improves significantly, compared to a simple benchmark scheme.

## I. INTRODUCTION

The forthcoming 5th Generation (5G) wireless networks will need to provide greatly improved spectral efficiency along with a defined quality of service (QoS) for real-time users (RTUs). A promising 5G technology is cell-free massive multiple-input multiple-output (MIMO), in which a large number of access points (APs) are randomly distributed through a coverage area and serve a much smaller number of users, providing uniform user experience [1]. The distributed APs are connected to a central processing unit (CPU) via high capacity backhaul links [1]–[6]. The problem of cell-free massive MIMO with limited backhal links has been considered in [4] and [7]. In [1], max-min fairness power control is exploited, while paper [2] studies the total energy efficiency optimization for cell-free massive MIMO taking into account the effect of backhaul power consumption. Different with previous work, in this paper, we investigate a mixed quality-of-service (QoS) problem [8]–[11] in which a set of RTUs requires a predefined rate and a max-min signal-to-interference-plus-noise ratio (SINR) is maintained between the non-real time users (NRTUs). The RTUs are defined as the users of real time services such as audio-video, video conferencing, web-based seminars, and video games, which result in the need for wireless communications with mixed QoS [8], [12]. A similar max–min SINR problem based on SINR known as *SINR balancing* in the literature has been considered in [13]–[15]. We show that the cell-free massive MIMO system has the capability of satisfying QoS requirements of the RTUs while it can guarantee excellent service for the NTTUs. The specific contributions of the paper are as follows:

1. An approximated SINR is derived based on the channel statistics and exploiting maximal ratio combining (MRC) at the APs. We formulate the corresponding mixed QoS problem with a fixed QoS requirement (i.e., SINR) for RTUs, which need to meet their target SINRs whereas the minimum SINRs of the remaining users should be maximized.

2. The mixed QoS problem is not jointly convex. We propose to deal with this non-convexity issue by decoupling the original problem into two sub-problems, namely, receiver filter coefficient design and power allocation.

3. It is shown that the receiver filter design problem can be solved through a generalized eigenvalue problem [16] whereas the user power allocation problem can be formulated using standard geometric programming (GP) [17]. An iterative algorithm is developed to solve the optimization problem. The convergence of the proposed scheme is explored numerically.

## II. SYSTEM MODEL

We consider uplink transmission in a cell-free massive MIMO system with $M$ randomly distributed APs and $K$ randomly distributed single-antenna users in the area. Moreover, we assume each AP has $N$ antennas. The channel coefficient vector between the $k$th user and the $m$th AP, $\mathbf{g}_{mk} \in \mathbb{C}^{N \times 1}$, is defined as $\mathbf{g}_{mk} = \sqrt{\beta_{mk}}\mathbf{h}_{mk}$, where $\beta_{mk}$ denotes the large-scale fading and $\mathbf{h}_{mk} \sim \mathcal{CN}(0,1)$ represents small-scale fading between the $k$th user and the $m$th AP [1]. All pilot sequences used in the channel estimation phase are collected in a matrix $\boldsymbol{\Phi} \in \mathbb{C}^{\tau \times K}$, where $\tau$ is the length of pilot sequence for each user and the $k$th column, $\boldsymbol{\phi}_k$, and represents the pilot sequence used for the $k$th user. The minimum mean square error (MMSE) estimate of the channel coefficient between the $k$th user and the $m$th AP is given by [1]

$$\hat{\mathbf{g}}_{mk} = c_{mk}\left(\sqrt{\tau p_p}\mathbf{g}_{mk} + \sqrt{\tau p_p}\sum_{k' \neq k}^{K}\mathbf{g}_{mk'}\boldsymbol{\phi}_{k'}^{H}\boldsymbol{\phi}_k + \mathbf{W}_{p,m}\boldsymbol{\phi}_k\right), \quad (1)$$

where each element of $\mathbf{W}_{p,m}$, $w_{p,m} \sim \mathcal{CN}(0,1)$, denotes the noise sequence at the $m$th antenna, $p_p$ represents the normalized signal-to-noise ratio (SNR) of each pilot symbol, and $c_{mk}$ is given by $c_{mk} = \frac{\sqrt{\tau p_p}\beta_{mk}}{\tau p_p \sum_{k'=1}^{K}\beta_{mk'}|\boldsymbol{\phi}_{k'}^{H}\boldsymbol{\phi}_k|^2 + 1}$. In this paper, we consider the uplink data transmission, where all users send their signals to the APs. The transmitted signal from the $k$th user is represented by $x_k = \sqrt{q_k}s_k$, where $s_k$ ($\mathbb{E}\{|s_k|^2\} = 1$) and $q_k$ denote the transmitted symbol and the transmit power at the $k$th user. The $N \times 1$ received signal at the $m$th AP from all users is given by $\mathbf{y}_m = \sqrt{\rho}\sum_{k=1}^{K}\mathbf{g}_{mk}\sqrt{q_k}s_k + \mathbf{n}_m$, where each element of $\mathbf{n}_m \in \mathbb{C}^{N \times 1}$, $n_{n,m} \sim \mathcal{CN}(0,1)$, is the noise at the $m$th AP and $\rho$ refers to the normalized SNR.

M. Bashar, K. Cumanan and A. G. Burr are with the Department of Electronic Engineering, University of York, UK. e-mail: {mb1465, kanapathippillai.cumanan, alister.burr}@york.ac.uk. H. Q. Ngo is with the School of Electronics, Electrical Engineering and Computer Science, Queen's University Belfast, UK. e-mail: hien.ngo@qub.ac.uk. H. Vincent Poor is with the Department of Electrical Engineering, Princeton University, Princeton, NJ, USA. e-mail:poor@princeton.edu.

The work of K. Cumanan and A. G. Burr was supported by H2020-MSC ARISE-2015 under Grant 690750.

## III. PERFORMANCE ANALYSIS

In this section, in deriving the achievable rate of each user, it is assumed that the CPU exploits only the knowledge of channel statistics between the users and APs in detecting data from the received signal in (2). The aggregated received signal at the CPU can be written as

$$r_k = \sum_{m=1}^{M} u_{mk} \left( \hat{\mathbf{g}}_{mk}^H \mathbf{y}_m \right). \quad (2)$$

By collecting all the coefficients $u_{mk}, \forall m$, corresponding to the $k$th user, we define $\mathbf{u}_k = [u_{1k}, u_{2k}, \cdots, u_{Mk}]^T$. To detect $s_k$, with the MRC processing, the aggregated received signal in (2) can be rewritten as

$$r_k = \sqrt{\rho}\mathbb{E}\left\{\sum_{m=1}^{M} u_{mk}\hat{\mathbf{g}}_{mk}^H \mathbf{g}_{mk}\sqrt{q_k}\right\} s_k \quad (3)$$
$$+ \sqrt{\rho}\left(\sum_{m=1}^{M} u_{mk}\hat{\mathbf{g}}_{mk}^H \mathbf{g}_{mk}\sqrt{q_k} - \mathbb{E}\left\{\sum_{m=1}^{M} u_{mk}\hat{\mathbf{g}}_{mk}^H \mathbf{g}_{mk}\sqrt{q_k}\right\}\right) s_k$$
$$\underbrace{}_{\text{BU}_k}$$
$$+ \sum_{k'\neq k}^{K} \underbrace{\sqrt{\rho}\sum_{m=1}^{M} u_{mk}\hat{\mathbf{g}}_{mk}^H \mathbf{g}_{mk'}\sqrt{q_{k'}}\,s_{k'}}_{\text{IUI}_{kk'}} + \underbrace{\sum_{m=1}^{M} u_{mk}\hat{\mathbf{g}}_{mk}^H \mathbf{n}_m}_{\text{TN}_k},$$

where $\text{DS}_k$ and $\text{BU}_k$ denote the desired signal (DS) and beamforming uncertainty (BU) for the $k$th user, respectively, and $\text{IUI}_{kk'}$ represents the inter-user-interference (IUI) caused by the $k'$th user. In addition, $\text{TN}_k$ accounts for the total noise (TN) following the MRC detection. The corresponding SINR can be defined by considering the worst-case of the uncorrelated Gaussian noise as follows [1]:

$$\text{SINR}_k = \frac{|\text{DS}_k|^2}{\mathbb{E}\{|\text{BU}_k|^2\} + \sum_{k'\neq k}^{K}\mathbb{E}\{|\text{IUI}_{kk'}|^2\} + \mathbb{E}\{|\text{TN}_k|^2\}}. \quad (4)$$

Based on the SINR definition in (4), the achievable uplink rate of the $k$th user is defined in the following theorem:

**Theorem 1.** *The achievable uplink rate of the $k$th user in the cell-free massive MIMO system with $K$ randomly distributed single-antenna users and $M$ APs is given by (5) (defined at the top of the next page).*

*Proof:* Please refer to Appendix A. ∎

Note that in (5), $\mathbf{u}_k = [u_{1k}, u_{2k}, \cdots, u_{Mk}]^T$, and the following equations hold: $\mathbf{\Gamma}_k = [\gamma_{1k}, \gamma_{2k}, \cdots, \gamma_{Mk}]^T$, $\gamma_{mk} = \sqrt{\tau_p p_p}\beta_{mk}c_{mk}$, $\mathbf{\Upsilon}_{kk'} = \text{diag}\left[\beta_{1k'}\gamma_{1k}, \cdots; \beta_{Mk'}\gamma_{Mk}\right]$, $\mathbf{\Lambda}_{kk'} = [\frac{\gamma_{1k}\beta_{1k'}}{\beta_{1k}}, \frac{\gamma_{2k}\beta_{2k'}}{\beta_{2k}}, \cdots, \frac{\gamma_{Mk}\beta_{Mk'}}{\beta_{Mk}}]^T$, and $\mathbf{R}_k = \text{diag}[\gamma_{1k}, \cdots, \gamma_{Mk}]$, and $\gamma_{mk} = \mathbb{E}\{|\hat{g}_{mk}|^2\} = \sqrt{\tau p_p}\beta_{mk}c_{mk}$.

## IV. PROPOSED MIXED QOS SCHEME

We formulate the mixed QoS problem, where the minimum uplink user rate among NRTUs is maximized while satisfying the transmit power constraint at each user and the RTUs' SINR target constraints. We assume users $1, 2, \cdots, K_1$ are RTUs. The mixed QoS problem is given by

$$P_1: \max_{q_k, \mathbf{u}_k} \min_{k=K_1+1,\cdots,K} R_k, \quad (6a)$$
$$\text{subject to} \quad 0 \leq q_k \leq p_{\max}^{(k)}, \quad \forall k, \quad (6b)$$
$$\text{SINR}_k^{\text{UP}} \geq \text{SINR}_k^t, k = 1, \cdots, K_1 \quad (6c)$$

where $p_{\max}^{(k)}$ is the maximum transmit power available at user $k$, and $\text{SINR}_k^t$ denotes the target SINR for the $k$th RTU. Problem $P_1$ is not jointly convex in terms of $\mathbf{u}_k$ and power allocation $q_k, \forall k$. Therefore, it cannot be directly solved through existing convex optimization software. To tackle this non-convexity issue, we decouple Problem $P_1$ into two sub-problems: receiver coefficient design (i.e. $\mathbf{u}_k$) and the power allocation problem, which are explained in the following subsections.

*1) Receiver Filter Coefficient Design:* In this subsection, the problem of designing the receiver coefficient is considered. These coefficients (i.e., $\mathbf{u}_k, \forall k$) are obtained by interdependently maximizing the uplink SINR of each user. Hence, the optimal receiver filter coefficients can be obtained through solving the following optimization problem:

$$P_2: \max_{\mathbf{u}_k} \frac{N^2 \mathbf{u}_k^H (q_k \mathbf{\Gamma}_k \mathbf{\Gamma}_k^H) \mathbf{u}_k}{\mathbf{u}_k^H \left( N^2 \sum_{k'\neq k}^{K} q_{k'}|\boldsymbol{\phi}_k^H \boldsymbol{\phi}_{k'}|^2 \mathbf{\Lambda}_{kk'} \mathbf{\Lambda}_{kk'}^H + N\sum_{k'=1}^{K} q_{k'} \mathbf{\Upsilon}_{kk'} + \frac{N}{\rho}\mathbf{R}_k \right) \mathbf{u}_k}. \quad (7)$$

Problem $P_2$ is a generalized eigenvalue problem [16], where the optimal solutions can be obtained by determining the generalized eigenvector of the matrix pair $\mathbf{A}_k = N^2 q_k \mathbf{\Gamma}_k \mathbf{\Gamma}_k^H$ and $\mathbf{B}_k = N^2 \sum_{k'\neq k}^{K} q_{k'}|\boldsymbol{\phi}_k^H \boldsymbol{\phi}_{k'}|^2 \mathbf{\Lambda}_{kk'} \mathbf{\Lambda}_{kk'}^H + N\sum_{k'=1}^{K} q_{k'} \mathbf{\Upsilon}_{kk'} + \frac{N}{\rho}\mathbf{R}_k$ corresponding to the maximum generalized eigenvalue.

*2) Power Allocation:* Next, we solve the power allocation problem for a given set of fixed receiver filter coefficients, $\mathbf{u}_k, \forall k$. The optimal transmit power can be determined by solving the following mixed QoS problem:

$$P_3: \max_{q_k} \min_{k=K_1+1,\cdots,K} \text{SINR}_k^{\text{UP}}, \quad (8a)$$
$$\text{subject to} \quad 0 \leq q_k \leq p_{\max}^{(k)}, \quad \forall k, \quad (8b)$$
$$\text{SINR}_k^{\text{UP}} \geq \text{SINR}_k^t. \; k = 1, \cdots, K_1 \quad (8c)$$

Note that the max-min rate problem and max-min SINR problem are equivalent. Without loss of generality, Problem $P_3$ can be rewritten by introducing a new slack variable as

$$P_4: \max_{t, q_k} t, \quad (9a)$$
$$\text{subject to} \quad 0 \leq q_k \leq p_{\max}^{(k)}, \quad \forall k, \quad (9b)$$
$$\text{SINR}_k^{\text{UP}} \geq t, \quad k = K_1+1, \cdots, K, \quad (9c)$$
$$\text{SINR}_k^{\text{UP}} \geq \text{SINR}_k^t, \quad k = 1, \cdots, K_1 \quad (9d)$$

**Proposition 1.** *Problem $P_4$ is a standard GP.*

*Proof:* The SINR constraint (9c) is not a posynomial functions in its form, however it can be rewritten into the following posynomial function:

$$\frac{\mathbf{u}_k^H \left( \sum_{k'\neq k}^{K} q_{k'}|\boldsymbol{\phi}_k^H \boldsymbol{\phi}_{k'}|^2 \mathbf{\Lambda}_{kk'} \mathbf{\Lambda}_{kk'}^H + \sum_{k'=1}^{K} q_{k'} \mathbf{\Upsilon}_{kk'} + \frac{1}{\rho}\mathbf{R}_k \right) \mathbf{u}_k}{\mathbf{u}_k^H \left( q_k \mathbf{\Gamma}_k \mathbf{\Gamma}_k^H \right) \mathbf{u}_k} < \frac{1}{t}. \quad (10)$$

$$R_k = \log_2 \left( 1 + \frac{\mathbf{u}_k^H \left( N^2 q_k \mathbf{\Gamma}_k \mathbf{\Gamma}_k^H \right) \mathbf{u}_k}{\mathbf{u}_k^H \left( N^2 \sum_{k' \neq k}^K q_{k'} |\boldsymbol{\phi}_k^H \boldsymbol{\phi}_{k'}|^2 \mathbf{\Lambda}_{kk'} \mathbf{\Lambda}_{kk'}^H + N \sum_{k'=1}^K q_{k'} \mathbf{\Upsilon}_{kk'} + \frac{N}{\rho} \mathbf{R}_k \right) \mathbf{u}_k} \right). \quad (5)$$

**Algorithm 1** Proposed algorithm to solve Problem $P_1$

1. Initialize $\mathbf{q}^{(0)} = [q_1^{(0)}, q_2^{(0)}, \cdots, q_K^{(0)}]$, $i = 1$
2. Repeat, $i = i + 1$
3. Set $\mathbf{q}^{(i)} = \mathbf{q}^{(i-1)}$ and determine the optimal receiver coefficients $\mathbf{U}^{(i)} = [\mathbf{u}_1^{(i)}, \mathbf{u}_2^{(i)}, \cdots, \mathbf{u}_K^{(i)}]$ through solving the generalized eigenvalue Problem $P_2$ in (7)
4. Compute $\mathbf{q}^{(i+1)}$ through solving Problem $P_4$ in (9)
5. Go back to Step 2 and repeat until required accuracy

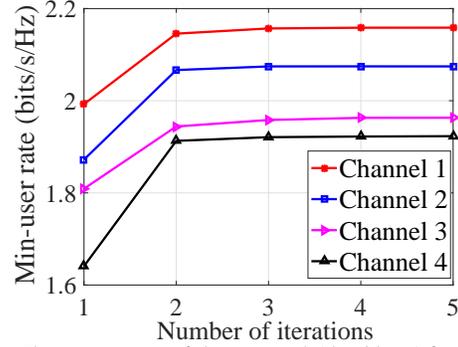

Figure 2. The convergence of the proposed Algorithm 1 for $M = 40$, $K = 22$, $K_1 = 2$, $N = 2$, $D = 1$ km, SINR$_k^t = 1$, and $\tau = 20$.

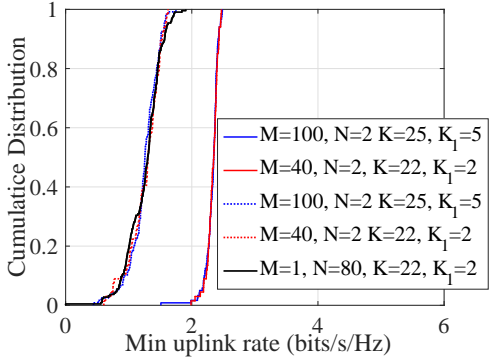

Figure 1. The cumulative distribution of the per-user uplink rate, for ($M = 100, K = 25, K_1 = 5$) and ($M = 40, K = 22, K_1 = 2$) with $D = 1$ km, $\tau = 20$, and SINR$_k^t = 1$. The solid curves refer to the proposed Algorithm 1, while the dashed curves present the case $u_{mk} = 1$, $\forall m, k$, and solve Problem $P_4$.

By applying a simple transformation, (10) can be rewritten in form of $q_k^{-1} \left( \sum_{k' \neq k}^K a_{kk'} q_{k'} + \sum_{k'=1}^K b_{kk'} q_{k'} + c_k \right) < \frac{1}{t}$, which shows that the left-hand side of (10) is a posynomial function. The same transformation holds for (9d). Therefore, Problem $P_4$ is a standard GP (convex problem). ∎

Based on two sub-problems, an iterative algorithm is developed by solving both sub-problems at each iteration. The proposed algorithm is summarized in Algorithm 1.

Moreover, investigating the effectiveness of the proposed scheme under realistic channel conditions [18], [19] is an interesting topic for future work.

## V. NUMERICAL RESULTS AND DISCUSSION

To model the channel coefficients between users and APs, the coefficient $\beta_{mk}$ is given by $\beta_{mk} = \text{PL}_{mk} . 10^{\frac{\sigma_{sh} z_{mk}}{10}}$ where $\text{PL}_{mk}$ is the path loss from the $k$th user to the $m$th AP, and $10^{\frac{\sigma_{sh} z_{mk}}{10}}$ denotes the shadow fading with standard deviation $\sigma_{sh}$, and $z_{mk} \sim \mathcal{N}(0, 1)$ [1]. The noise power is given by $P_n = \text{BW} k_B T_0 W$, where BW = 20 MHz denotes the bandwidth, $k_B = 1.381 \times 10^{-23}$ represents the Boltzmann constant, and $T_0 = 290$ (Kelvin) denotes the noise temperature. Moreover, $W = 9$dB, and denotes the noise figure [1]. It is assumed that that $\bar{P}_p$ and $\bar{\rho}$ denote the transmit powers of the pilot and data symbols, respectively, where $P_p = \frac{\bar{P}_p}{P_n}$ and $\rho = \frac{\bar{\rho}}{P_n}$. In simulations, we set $\bar{P}_p = 200$ mW and $\bar{\rho} = 200$ mW.

A cell-free massive MIMO system is considered with 15 APs ($M = 15$) and 6 users ($K = 6$) who are randomly distributed over the coverage area of size $1 \times 1$ km. Moreover, each AP is equipped with $N = 3$ antennas and we set the total number of RTUs to $K_1 = 2$, and random pilot sequences with length $\tau = 5$ are considered. Table I presents the achievable SINRs of the users while the target SINR for both RTUs is fixed as 2.3. The power allocations for all users and the max-min SINR values are obtained using the proposed Algorithm 1. It can be seen from Table I that both RTUs achieve their target SINR, while the minimum SINR of the rest of the users is maximized through using Algorithm 1 (If the problem is infeasible, we set SINR$_k = 0, \forall k$). Fig. 1 presents the cumulative distribution of the achievable uplink rates for the proposed Algorithm 1 (the solid curves) and a scheme in which the received signals are not weighted (i.e. we set $u_{mk} = 1$, $\forall m, k$ and solve Problem $P_4$), which are shown by the dashed curves. As seen in Fig. 1, the median of the cumulative distribution of the minimum uplink rate of the users is significantly increased compared to the scheme with $u_{mk} = 1, \forall m, k$ and solving Problem $P_4$. As seen in Fig. 1, the performance (i.e. the 10% outage rate) of the proposed scheme is almost twice that of the case with $u_{mk} = 1 \forall m, k$. Note that the authors in [1] considered a max-min SINR problem defining only power coefficients and without QoS constraints for RTUs. Hence, the dashed curves in Fig. 1 refer to the scheme in [1] along with QoS constraints. Moreover, note that the case with $M = 1$ and $N = 80$ refers to the single-cell massive MIMO system, in which all service antennas are collocated at the center of cell. As the figure demonstrates the performance of cell-free massive MIMO is significantly better than the conventional single-cell massive MIMO system. Fig. 2 demonstrates numerically the convergence of the proposed Algorithm 1 with 20 APs ($M = 20$) and 20 users ($K = 20$) and random pilot sequences with length $\tau = 15$. At each iteration, one of the design parameters is determined by solving the corresponding sub-problem while other design variables are fixed. Assume that at the $i$th iteration, the receiver filter coefficients $\mathbf{u}_k^{(i)}$, $\forall k$ are determined for a fixed power allocation $\mathbf{q}^{(i)}$ and the power allocation $\mathbf{q}^{(i+1)}$ is obtained for a given set of receiver filter coefficients $\mathbf{u}_k^{(i)}$, $\forall k$. The optimal power allocation $\mathbf{q}^{(i+1)}$ obtained for a given $\mathbf{u}_k^{(i)}$ achieves an uplink rate greater than or equal to that of the previous iteration. As a result, the achievable uplink rate monotonically increases at each iteration, which can be also observed from the numerical results presented in Fig. 2.

Table I
TARGET SINRS AND THE POWER CONSUMPTION OF THE PROPOSED SCHEME, WITH $M = 15$, $N = 3$, $K = 6$, $K_1 = 2$, $\tau = 5$, AND $D = 1$ KM.

| | Achieved SINR | | | | | | Power Allocation ($q_k$) | | | | | |
|---|---|---|---|---|---|---|---|---|---|---|---|---|
| Channels | RTU1 | RTU2 | NRTU1 | NRTU2 | NRTU3 | NRTU4 | RTU1 | RTU2 | NRTU1 | NRTU2 | NRTU3 | NRTU4 |
| Channel 1 | 2.3 | 2.3 | 0.6457 | 0.6457 | 0.6457 | 0.6457 | 0.0519 | 0.1472 | 0.2039 | 0.3111 | 0.0056 | 1 |
| Channel 2 | 2.3 | 2.3 | 0.7445 | 0.7445 | 0.7445 | 0.7445 | 0.2995 | 0.0098 | 0.0050 | 1 | 0.3398 | 0.2278 |
| Channel 3 | 2.3 | 2.3 | 0.6479 | 0.6479 | 0.6479 | 0.6479 | 0.7001 | 0.1045 | 0.0085 | 0.0170 | 1 | 0.1415 |
| Channel 4 | 2.3 | 2.3 | 1.9622 | 1.9622 | 1.9622 | 1.9622 | 0.0296 | 0.0438 | 1 | 0.1753 | 0.0379 | 0.4827 |

## VI. CONCLUSIONS

We have investigated the mixed QoS problem with QoS requirements for the RTUs in cell-free massive MIMO, and proposed a solution to maximize the minimum user rate while satisfying the SINR constraints of the RTUs. To realize the solution, the original mixed QoS problem has been divided into two sub-problems and they have been iteratively solved by formulating them into a generalized eigenvalue problem and GP.

## APPENDIX A: PROOF OF THEOREM 1

The desired signal for the user $k$ is given by $\mathrm{DS}_k = \sqrt{\rho}\mathbb{E}\left\{\sum_{m=1}^{M} u_{mk}\hat{\mathbf{g}}_{mk}^H \mathbf{g}_{mk}\sqrt{q_k}\right\} = N\sqrt{\rho q_k}\sum_{m=1}^{M} u_{mk}\gamma_{mk}$. Hence, $|\mathrm{DS}_k|^2 = \rho q_k \left(N \sum_{m=1}^{M} u_{mk}\gamma_{mk}\right)^2$. Moreover, the term $\mathbb{E}\{|\mathrm{BU}_k|^2\}$ can be obtained as

$$\mathbb{E}\left\{|\mathrm{BU}_k|^2\right\} = \rho\mathbb{E}\left\{\left|\sum_{m=1}^{M} u_{mk}\hat{\mathbf{g}}_{mk}^H \mathbf{g}_{mk}\sqrt{q_k}\right.\right. \tag{11}$$

$$\left.\left. - \rho\mathbb{E}\left\{\sum_{m=1}^{M} u_{mk}\hat{\mathbf{g}}_{mk}^H \mathbf{g}_{mk}\sqrt{q_k}\right\}\right|^2\right\} = \rho N \sum_{m=1}^{M} q_k u_{mk}^2 \gamma_{mk}\beta_{mk},$$

where the last equality comes from the analysis in [1, Appendix A]. The term $\mathbb{E}\{|\mathrm{IUI}_{kk'}|^2\}$ is obtained as

$$\mathbb{E}\{|\mathrm{IUI}_{kk'}|^2\} = \rho\mathbb{E}\left\{\left|\sum_{m=1}^{M} u_{mk}\hat{\mathbf{g}}_{mk}^H \mathbf{g}_{mk'}\sqrt{q_{k'}}\right|^2\right\}$$

$$= \rho q_{k'}\mathbb{E}\left\{\left|\underbrace{\sum_{m=1}^{M} c_{mk}u_{mk}\mathbf{g}_{mk'}^H \tilde{\mathbf{w}}_{mk}}_{A}\right|^2\right\}$$

$$+ \rho\tau p_p\mathbb{E}\left\{q_{k'}\left|\underbrace{\sum_{m=1}^{M} c_{mk}u_{mk}\left(\sum_{i=1}^{K} \mathbf{g}_{mi}\boldsymbol{\phi}_k^H \boldsymbol{\phi}_i\right)^H \mathbf{g}_{mk'}}_{B}\right|^2\right\}. \tag{12}$$

Since $\tilde{\mathbf{w}}_{mk} = \boldsymbol{\phi}_k^H \mathbf{W}_{\mathbf{p},\mathbf{m}}$ is independent from the term $g_{mk'}$ similar to [1, Appendix A], the term $A$ in (12) immediately is given by $A = Nq_{k'}\sum_{m=1}^{M} c_{mk}^2 u_{mk}^2 \beta_{mk'}$. The term $B$ in (12) can be obtained as

$$B = \tau p_p q_{k'}\mathbb{E}\left\{\underbrace{\left|\sum_{m=1}^{M} c_{mk}u_{mk}\|\mathbf{g}_{mk'}\|^2 \boldsymbol{\phi}_k^H \boldsymbol{\phi}_{k'}\right|^2}_{C}\right\}$$

$$+ \tau p_p q_{k'}\mathbb{E}\left\{\underbrace{\left|\sum_{m=1}^{M} c_{mk}u_{mk}\left(\sum_{i\neq k'}^{K} \mathbf{g}_{mi}\boldsymbol{\phi}_k^H \boldsymbol{\phi}_i\right)^H \mathbf{g}_{mk'}\right|^2}_{D}\right\}. \tag{13}$$

The first term in (13) is given by

$$C = N\tau p_p q_{k'}\left|\boldsymbol{\phi}_k^H \boldsymbol{\phi}_{k'}\right|^2 \sum_{m=1}^{M} c_{mk}^2 u_{mk}^2 \beta_{mk'}^2$$

$$+ N^2 q_{k'}\left|\boldsymbol{\phi}_k^H \boldsymbol{\phi}_{k'}\right|^2 \left(\sum_{m=1}^{M} u_{mk}\gamma_{mk}\frac{\beta_{mk'}}{\beta_{mk}}\right)^2, \tag{14}$$

where the last equality is derived based on the fact $\gamma_{mk} = \sqrt{\tau p_p}\beta_{mk}c_{mk}$. The second term in (13) can be obtained as

$$D = N\sqrt{\tau p_p}q_{k'}\sum_{m=1}^{M} u_{mk}^2 c_{mk}\beta_{mk'}\beta_{mk} - Nq_{k'}\sum_{m=1}^{M} u_{mk}^2 c_{mk}^2 \beta_{mk'}$$

$$- N\tau p_p q_{k'}\sum_{m=1}^{M} u_{mk}^2 c_{mk}^2 \beta_{mk'}^2 \left|\boldsymbol{\phi}_k^H \boldsymbol{\phi}_{k'}\right|^2. \tag{15}$$

Finally we obtain

$$\mathbb{E}\{|\mathrm{IUI}_{kk'}|^2\} = N\rho q_{k'}\left(\sum_{m=1}^{M} u_{mk}^2 \beta_{mk'}\gamma_{mk}\right)$$

$$+ N^2 \rho q_{k'}\left|\boldsymbol{\phi}_k^H \boldsymbol{\phi}_{k'}\right|^2 \left(\sum_{m=1}^{M} u_{mk}\gamma_{mk}\frac{\beta_{mk'}}{\beta_{mk}}\right)^2. \tag{16}$$

The total noise for the user $k$ is given by $\mathbb{E}\{|\mathrm{TN}_k|^2\} = \mathbb{E}\left\{\left|\sum_{m=1}^{M} u_{mk}\hat{\mathbf{g}}_{mk}^H \mathbf{n}_m\right|^2\right\} = N\sum_{m=1}^{M} u_{mk}^2 \gamma_{mk}$. Finally, SINR of user $k$ is obtained by (5), which completes the proof. ∎